\documentclass[aps,pra,twocolumn,amsmath,superscriptaddress,longbibliography]{revtex4-2}

\usepackage{amsmath}
\usepackage[urlcolor=blue,colorlinks=true,citecolor=blue,linkcolor=blue,pdfstartview={FitH},bookmarks=false]{hyperref}
\usepackage{appendix}
\usepackage{graphicx}
\usepackage{longtable}
\usepackage{epsfig}
\usepackage{dcolumn}
\usepackage{bm}
\usepackage{bbm}
\usepackage{amssymb}
\usepackage{multirow}
\usepackage{times,color}
\usepackage{hyperref}
\usepackage{amsmath}
\usepackage{color}
\usepackage{subfigure}
\usepackage{float}
\usepackage{epstopdf}

\begin{document}

\title{Unveiling quantum criticality of disordered Aubry-Andr\'{e}-Harper models via typical fidelity susceptibility}
\author {Tian-Cheng Yi}

\affiliation{Department of Physics and Zhejiang Key Laboratory of Quantum State Control and Optical Field Manipulation,
Zhejiang Sci-Tech University, Hangzhou 310018, China}

\author {Ying-Ying Fang}
\affiliation{Department of Physics and Zhejiang Key Laboratory of Quantum State Control and Optical Field Manipulation,
Zhejiang Sci-Tech University, Hangzhou 310018, China}

\author{Wen Chen}

\affiliation{Department of Physics, University of Houston, Houston, Texas 77004, USA}
\author{Wen-Long You}
\email{wlyou@nuaa.edu.cn}
\affiliation{College of Physics, Nanjing University of Aeronautics and Astronautics, Nanjing, 211106, China}
\author {Yunbo Zhang}
\email{ybzhang@zstu.edu.cn}
\affiliation{Department of Physics and Zhejiang Key Laboratory of Quantum State Control and Optical Field Manipulation,
Zhejiang Sci-Tech University, Hangzhou 310018, China}

\begin{abstract}
In this study, we investigate the localization transition and quantum criticality {in the ground state of the} disordered Aubry-Andr\'{e}-Harper (AAH) model, where a quasiperiodic potential is hybridized with a disordered potential.
In the clean limit, the AAH model undergoes a localization transition from an extended phase to a localized phase via an intermediate critical phase as the strength of the quasiperiodic potential is varied. 
While the staggered potential merely shifts the critical point to a lower value, Fibonacci and Thue-Morse potentials induce immediate localization. This contrast reveals the sensitivity of localization behavior to the structural complexity of the potential, with the onset of localization correlating with the sequence's complexity. More specifically, the system follows a hierarchy defined by the complexity measures of the applied potentials. 
In addition, the typical fidelity susceptibility exhibits a power-law scaling behavior at the localization transition, enabling reliable extraction of the critical exponent. 
We focus on the AAH model with the Fibonacci potential due to its minimal finite-size effects compared to other cases. 
For the disordered AAH model with the Fibonacci potential, we determine critical exponents that differ from those of the AAH model without disorder and the Anderson model.
Moreover, despite differences in localization behavior, we find that the disordered AAH models with the staggered potential and the Fibonacci potential share the same correlation-length critical exponent.
These findings provide a unified framework for understanding localization transitions in quasiperiodic systems and are amenable to experimental validation using emerging techniques.
\end{abstract}
\maketitle
\section{Introduction}
\label{sec1}
The study of hybrid potentials that combines quasiperiodicity~\cite{harper1955single, aubry1980analyticity,PhysRevB.23.6422,PhysRevLett.61.2144,RevModPhys.65.213,PhysRevA.80.021603,PhysRevB.83.075105,PhysRevLett.108.220401,PhysRevLett.110.076403,PhysRevLett.110.180403,PhysRevLett.120.160404,PhysRevB.97.104202,PhysRevA.105.013315,PhysRevB.106.144205,PhysRevA.108.033305,PhysRevE.109.054123}  and disorder~\cite{PhysRev.109.1492,RevModPhys.80.1355,PhysRevA.99.042117,PhysRevB.104.115414,PhysRevB.104.195117,PhysRevLett.128.116801} is essential for advancing our understanding of complex quantum systems beyond ideal models. Real materials in nature often possess potential fields that simultaneously exhibit ordered and disordered characteristics, which cannot be fully described by purely periodic, quasiperiodic, or completely disordered models. While numerous previous studies have extensively explored mixed effects of quasiperiodic-quasiperiodic hybrids~\cite{xiaoming2022equivalence,lesser2022emergence,gonifmmode2023renormalization,chakrabarti2024strongly,PhysRevLett.126.106803,PhysRevB.105.214203,PhysRevB.107.035402} and quasiperiodic-periodic hybrids~\cite{padhan2022emergence,qi2023multiple,PhysRevB.111.134204}, the hybrid mechanisms between quasiperiodic and disordered potentials remain unclear.

A typical model of quasiperiodic systems is the Aubry-Andr\'{e}-Harper (AAH) model~\cite{harper1955single, aubry1980analyticity}, where quasiperiodicity is introduced specifically through a cosine modulation incommensurate with the lattice spacing. Existing research on quasiperiodic-disordered hybrid systems has primarily focused on the combination of the AAH model and Anderson disorder~\cite{bu2022quantum, bu2023kibble,PhysRevB.111.024205,PhysRevB.110.024207}, where the excessive degrees of freedom in Anderson disorder complicate both theoretical analysis and experimental control. In contrast, discrete quasi-disordered systems, such as Fibonacci~\cite{PhysRevB.35.1020, PhysRevB.37.5723, PhysRevE.99.032415, PhysRevE.102.012104, chenwen2024Multifractality, jagannathan2021theFibonacci} or Thue-Morse chains~\cite{PhysRevB.37.4375, PhysRevB.43.1034, PhysRevB.46.5162}, demonstrate intermediate properties between order and disorder: they retain long-range correlations characteristic of quasiperiodicity while preserving features of the Anderson model.

Discrete disorder (such as Fibonacci or Thue-Morse sequences) provides a unique platform bridging order and randomness. Unlike fully random Anderson disorder, these deterministic sequences possess well-defined structural correlations while introducing aperiodicity, making them experimentally feasible for quantum simulators (e.g., optical lattices~\cite{PhysRevA.92.063426, PhysRevB.110.104513} or quantum walks~\cite{PhysRevLett.131.150803, PhysRevA.109.012409, PhysRevA.110.052410}). Additionally, such different types of discrete potentials can be easily compared in terms of their complexity.

In this work, we investigate {the ground-state properties of} the disordered AAH model with staggered, Fibonacci, and Thue-Morse potentials.
In the absence of discrete potential, the AAH model exhibits extended, critical, and localized phases. The localization transition progresses from the extended phase to the critical phase and ultimately to the localized phase as the AAH potential is tuned. Without a disorder potential, the system features a self-dual point. When the staggered potential is introduced, the critical point shifts to a smaller value. In contrast, when the AAH potential coexists with either the Fibonacci or Thue-Morse potential, the system remains in a localized phase. 
In addition, the typical fidelity susceptibility exhibits a power-law scaling behavior at the localization transition, enabling reliable extraction of the critical exponent. 
We focus on the AAH model with the Fibonacci potential due to its minimal finite-size effects compared to other cases. 
For the disordered AAH model with the Fibonacci potential, we identify critical exponents that differ from those of the AAH model without disorder and the Anderson model. 
Furthermore, despite differences in localization behavior, we find that the disordered AAH models with the staggered potential and the Fibonacci potential share the same correlation-length critical exponent. 

The organization of the paper is as follows. In Sec.~\ref{sec2}, we introduce the Hamiltonian of the AAH model with staggered, Fibonacci, and Thue-Morse potentials. In Sec.~\ref{sec3}, we present the phase diagram of the AAH model with various discrete potentials, using the IPR as a tool. We observe that the amplitude of the IPR in these hybrid models partially follows the hierarchy established by Lempel-Ziv complexity measures. However, this pattern is disrupted when the AAH potential becomes dominant. In Sec.~\ref{sec4}, we investigate the quantum criticality of the disordered AAH model by employing the typical fidelity susceptibility. Finally, conclusions are presented in Sec.~\ref{sec5}.

\section{Model}
\label{sec2}
\begin{figure}[!ht]
\centering
\includegraphics[width=0.45\textwidth]{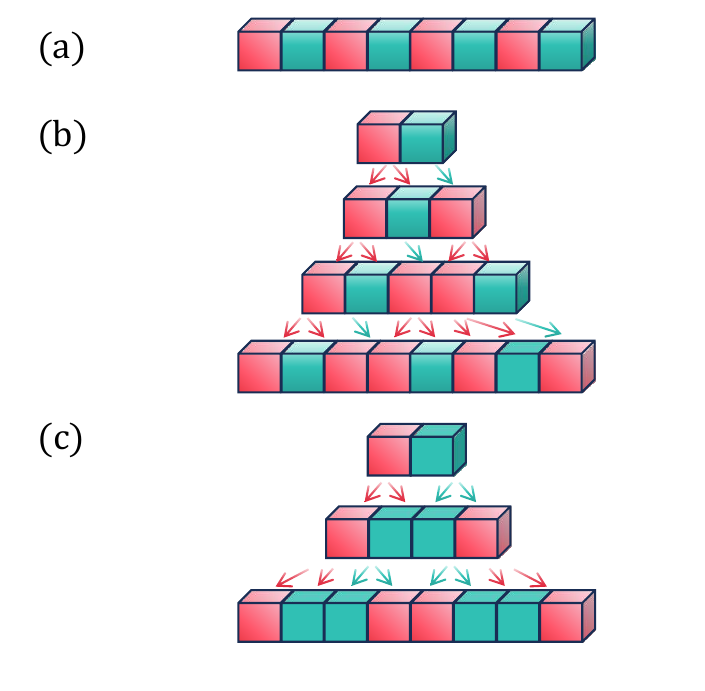}
\caption{
Schematic of (a) staggered, (b) Fibonacci, and (c) Thue-Morse sequences, with generation rules shown for (b) and (c).
Red cubes denote $+1$ and blue cubes denote $-1$. For the Fibonacci potential, the generation rule is $(1 \rightarrow 1, -1), (-1 \rightarrow 1)$ . For the Thue-Morse potential, the generation rule is $(1 \rightarrow 1, -1), (-1 \rightarrow -1, 1)$.
}   
\label{Schematic}
\end{figure}

The disordered AAH model with a discrete potential is given by the following Hamiltonian:
\begin{eqnarray}
\hat{H}&=&-t\sum_{i=1}^N\left(\hat{c}_{i}^{\dagger}\hat{c}_{i+1}+{\rm H.c.}\right)\nonumber\\
&&+V\sum_{i=1}^N\cos(2\pi\alpha i+\varphi) \hat{c}_{i}^{\dagger}\hat{c}_{i} 
+\Delta\sum_{i=1}^N w_i \hat{c}_{i}^{\dagger}\hat{c}_{i},
\label{ham}
\end{eqnarray}
where $c_{i}^{\dag}$ ($c_{i}$) is the fermionic creation (annihilation) operator at the $i$-th site among a total of $N$ lattice sites. We set the nearest-neighbor hopping strength $t = 1$ as the energy unit and H.c. represents the Hermitian conjugate. One of the external potentials in Eq.~(\ref{ham}) is the quasiperiodic term $V\cos(2\pi\alpha i + \varphi)$, where $V$ is the strength of the incommensurate potential, $\alpha$ is the irrational number $({\sqrt{5} - 1})/{2}$ (the inverse golden ratio) and $\varphi \in [0, 2\pi)$ is a random phase. In numerical calculations, the golden ratio $\alpha$ is approximated by rational numbers $\alpha = {F_k}/{F_{k+1}}$, where $F_k$ is the $k$-th Fibonacci number~\cite{SM}.
In this work, we employ periodic boundary conditions.
Another external potential is represented by $w_i$, defined as a combination of $+1$ and $-1$ following specific rules. Examples include staggered potentials or deterministic sequences such as the Fibonacci, and Thue-Morse sequences. Schematic diagrams illustrating the staggered, Fibonacci, and Thue-Morse sequences are shown in Fig.~\ref{Schematic}, where red cubes denote $+1$ and blue cubes denote $-1$. For the Fibonacci potential, the generation rule is $(1 \rightarrow 1, -1), (-1 \rightarrow 1)$ [see Fig.~\ref{Schematic}(b)]. For the Thue-Morse potential, the generation rule is $(1 \rightarrow 1, -1), (-1 \rightarrow -1, 1)$ [cf Fig.~\ref{Schematic}(c)].
$\Delta$ denotes the amplitude of such discrete disorder.
{In this work, we focus on the ground-state properties of the disordered AAH model with a discrete potential.}
\begin{figure}[!t]
\centering
\includegraphics[width=0.45\textwidth]{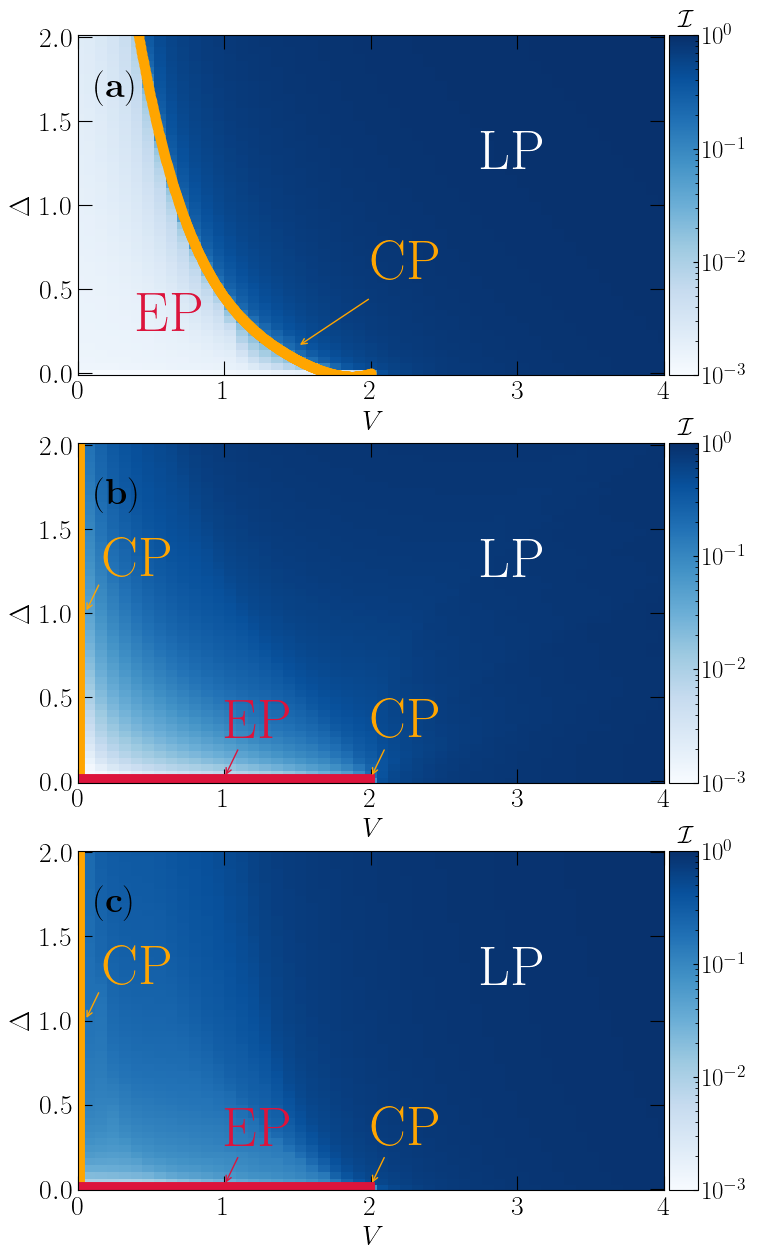}
\caption{
The {ground-state} phase diagrams of the disordered AAH models~(\ref{ham}) with (a) staggered potential, (b) Fibonacci potential, and (c) Thue-Morse potential for $N = 987$. 
The colormap represents the IPR ${\cal I}$. 
For $\Delta=0$,
When $V<2$, the ground state is in the extended phase (EP);
When $V=2$, the ground state is in the critical phase (CP);
When $V>2$, the ground state is in the localized phase (LP).
For $\Delta>0$, the staggered
potential merely shifts the critical point to a lower value, while Fibonacci and Thue-Morse potentials induce immediate
localization.
In (b) and (c), the red line corresponds to the EP for $\Delta = 0$, and the yellow line marks the CP for $V = 0$.
}   
\label{phasef}
\end{figure}
\begin{figure}[!t]
\centering
\includegraphics[width=0.45\textwidth]{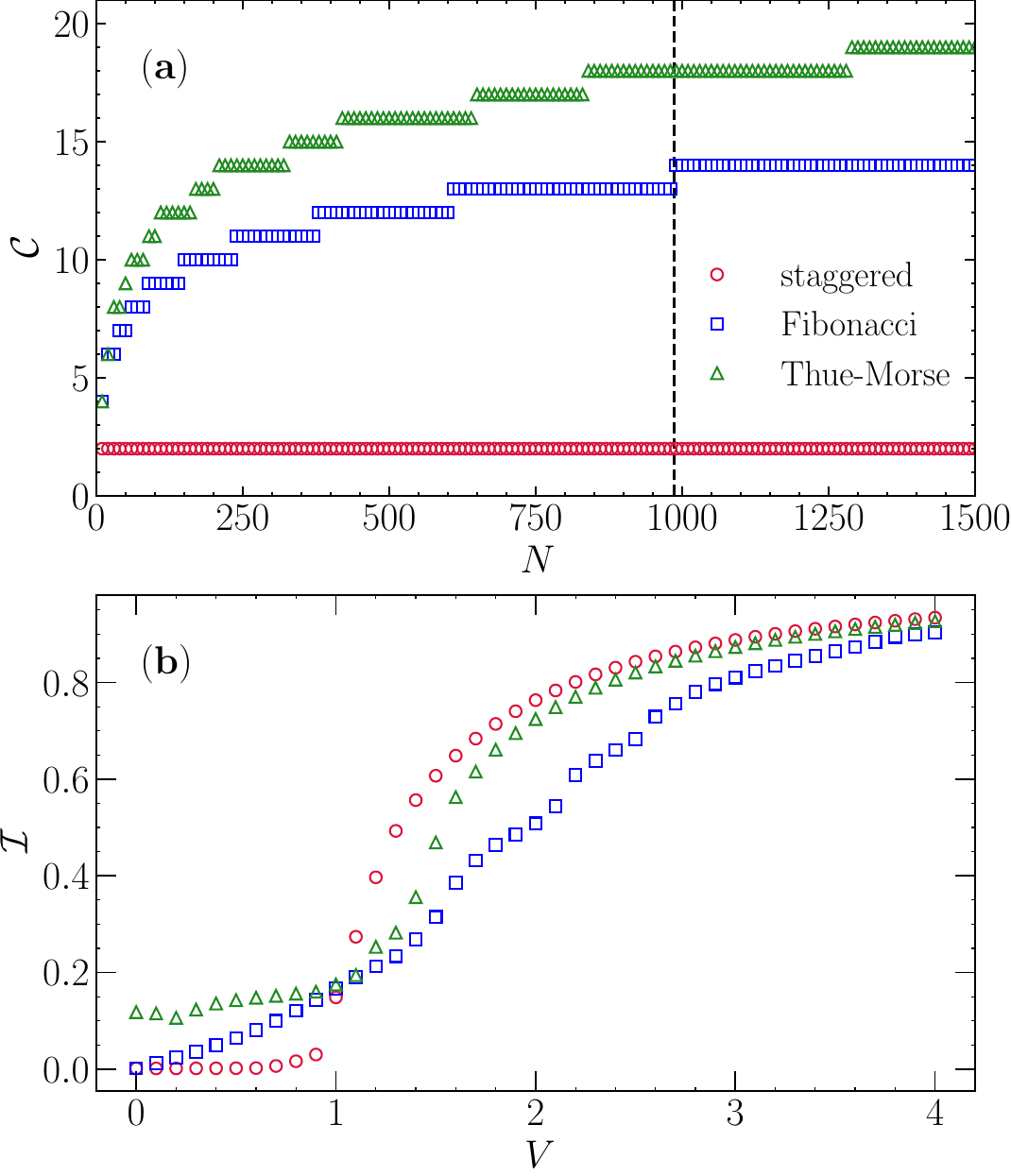}
\caption{
The correlations between the Lempel-Ziv complexity ${\cal C}$ and the IPR ${\cal I}$: (a) The Lempel-Ziv complexity ${\cal C}$ for the staggered, Fibonacci, and Thue-Morse potentials as a function of the system size $N$. The dashed line marks $N = 987$. (b) The IPR ${\cal I}$ as a function of $V$ for the AAH model with staggered, Fibonacci, and Thue-Morse potentials for $N = 987$ when $\Delta = 0.5$.
}   
\label{LZC}
\end{figure}

\section{Phase diagram and Lempel-Ziv complexity}
\label{sec3}

The eigenspectra and eigenstates of Eq.~(\ref{ham}) can be determined by solving the Schrödinger equation
\begin{eqnarray}
\hat{H}\left|\Psi_n\right\rangle = E_n\left|\Psi_n\right\rangle,
\end{eqnarray}
where $\left|\Psi_n\right\rangle = \sum_j \psi_{n, j}|j\rangle$ is expressed in terms of the Wannier basis $|j\rangle$, representing the state of a single particle localized at the $j$-th site of the lattice. The Hamiltonian (\ref{ham}) can be recast as
\begin{eqnarray}
\hat{H} = \hat{\mathbf{c}}^{\dagger} \mathcal{H} \hat{\mathbf{c}},
\end{eqnarray}
where $\hat{\mathbf{c}} = \left[\hat{c}_1, \ldots, \hat{c}_{N}\right]^T$ and $\mathcal{H}$ is an $N \times N$ single-particle Hamiltonian matrix, given by
\begin{eqnarray}
\mathcal{H} = \left(
\begin{array}{ccccccc}
\varepsilon_1 & -t & 0 & \cdots & \cdots & \cdots & -t \\
-t & \varepsilon_2 & -t & 0 & \cdots & \cdots & 0 \\
0 & -t & \varepsilon_3 & -t & \ddots & \ddots & 0 \\
\vdots & \ddots & \ddots & \ddots & \ddots & \ddots & \vdots \\
0 & \cdots & 0 & -t & \varepsilon_{N-2} & -t & 0 \\
0 & \cdots & \cdots & 0 & -t & \varepsilon_{N-1} & -t \\
-t & \cdots & \cdots & \cdots & \cdots & -t & \varepsilon_{N}
\end{array}
\right),
\end{eqnarray}
where $\varepsilon_i = V\cos(2\pi\alpha i + \varphi) + \Delta w_i$.
By diagonalizing the Hamiltonian~(\ref{ham}), one can obtain the wave function $\Psi_n$.

To distinguish between spatially localized and extended states across all energy levels, we calculate the normalized inverse participation ratio (IPR) as a quantitative measure for the $n$-th eigenstate $|\Psi_n\rangle$ of the model (\ref{ham}), given by~\cite{misguich2016Iinverse, licciardello1978conductivity}  
\begin{eqnarray}
{\cal I} = \frac{\sum_{j=1}^N |\psi_{n,j}|^4}{\left( \sum_{j=1}^N |\psi_{n,j}^2| \right)^2}.
\label{equ:IPR}
\end{eqnarray}
In disordered systems, the IPR serves as a diagnostic tool to distinguish between localized and extended quantum states. For extended states, the IPR scales as $\mathcal{I} \propto N^{-1}$, while for localized states, $\mathcal{I} \propto N^0$. Critical states exhibit fractal characteristics, with scaling $\mathcal{I} \propto N^{-{\cal D}}$, where $0 < {\cal D} < 1$ represents the fractal dimension~\cite{PhysRevA.109.023314, PhysRevB.110.184207, PhysRevB.111.134204}.

In Figs.~\ref{phasef}(a-c), we present the {ground-state} phase diagrams of the disordered AAH model~(\ref{ham}) with staggered, Fibonacci, and Thue-Morse potentials for $N=987$. The phase diagrams feature extended, critical, and localized phases, with the colormap representing the IPR ${\cal I}$. In the absence of discrete disorder, the AAH model supports extended, critical, and localized phases. The localization transition occurs as the AAH potential is tuned, progressing from the extended phase to the critical point, and finally to the localized phase. Without a disorder potential, the system exhibits a self-dual point. When the staggered potential is introduced, the critical point shifts to a smaller value, as shown in Fig.~\ref{phasef}(a). In Figs.~\ref{phasef}(b-c), when $V = 0$ and $\Delta \neq 0$, the disordered AAH model~(\ref{ham}) reduces to the Fibonacci model and the Thue-Morse model, respectively, with the corresponding ground states lying in the critical phase~\cite{jagannathan2021theFibonacci, PhysRevB.37.4375, PhysRevB.43.1034, PhysRevB.46.5162}. When both $\Delta > 0$ and $V > 0$, corresponding to the coexistence of the AAH potential with either the Fibonacci or Thue-Morse potential, the system is always in a localized phase.

Next, we explore the relationships between various disordered models using the Lempel-Ziv complexity, ${\cal C}$~\cite{PhysRevB.104.195117}.
The Lempel-Ziv complexity is a concept from computer science that quantifies the complexity of a sequence. It is based on the idea of compressibility, specifically the amount of information required to describe a sequence without redundancy. The Lempel-Ziv complexity is calculated through the following steps: (1) initialize an empty dictionary to store existing phrases; (2) iteratively parse the sequence, finding the longest match in the dictionary, appending the next character, and updating the dictionary; (3) count the total number of phrases generated. For example, the string "ABAABABA" can be parsed into the following phrases: A, B, AA, BA, yielding ${\cal C} = 4$.

In Fig.~\ref{LZC}(a), we show the complexity of the different {discrete} potentials. The sequence complexities, in ascending order, are: staggered, Fibonacci, and Thue-Morse. In Fig.~\ref{LZC}(b), we present the IPR as a function of $V$ for  the disordered AAH model with various discrete potentials, with the system size $N = 987$ and $\Delta = 0.5$. {For the AAH model with staggered potential}, the IPR is finite when $V > 0.9$, indicating a phase transition at $V = 0.9$. The disordered AAH model with Fibonacci or Thue-Morse potential exhibits different behavior, where any non-zero finite discrete potential leads to an increase in IPR, demonstrating the system's localization (more related results iis shown in \cite{SM}). 
Furthermore, the order of the IPR follows the order of the complexity {when the amplitude of the AAH potential is relatively weak}. {However, as $V$ increases, the AAH potential becomes dominant. Since the complexity characterizes the nature of the discrete potential, this correspondence breaks down in that regime.}
If the complexity is constant (i.e., does not vary with size), such discrete potentials can lead to a phase transition when $V$ is finite. However, when the complexity increases with the system size, any finite discrete disorder will lead to localization. This rule also applies to other discrete potentials $\Delta$~\cite{SM}.

\begin{figure}[!htb]
\centering
\includegraphics[width=0.45\textwidth]{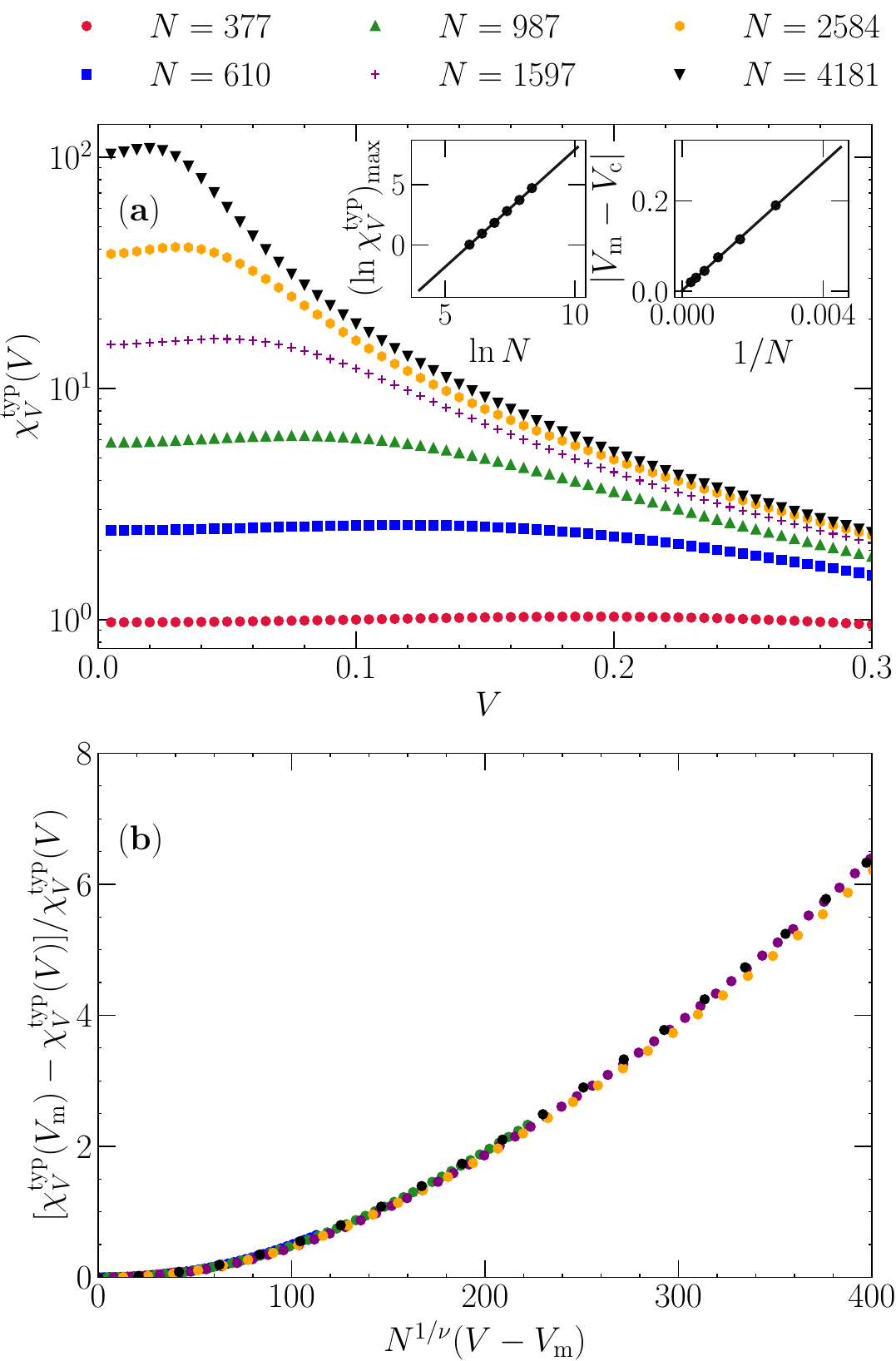} 
\caption{
Scaling properties of the disordered AAH model with Fibonacci potential for $\Delta = 0.1$: (a) The quantum fidelity susceptibility $ \chi_V^{\rm typ}$ as a function of the AAH potential strength $V$ for different system sizes. The inset shows the log-log scaling of $(\ln \chi_V^{\rm typ})_{\rm max}$ versus system size $N$ and the absolute value of the difference between the pseudocritical point and the critical point $|V_{\rm m} - V_{\rm c}|$ versus $1/N$. (b) The rescaled fidelity susceptibility, $[\chi_V^{\rm typ}(V_{\rm m}) - \chi_V^{\rm typ}(V)] / \chi_V^{\rm typ}(V)$, plotted as a function of the scaling variable $N^{1/\nu}(V - V_{\rm m})$ for $\nu = 1$.
}
\label{fsfv}
\end{figure}

\begin{figure}[!htb]
\centering
\includegraphics[width=0.45\textwidth]{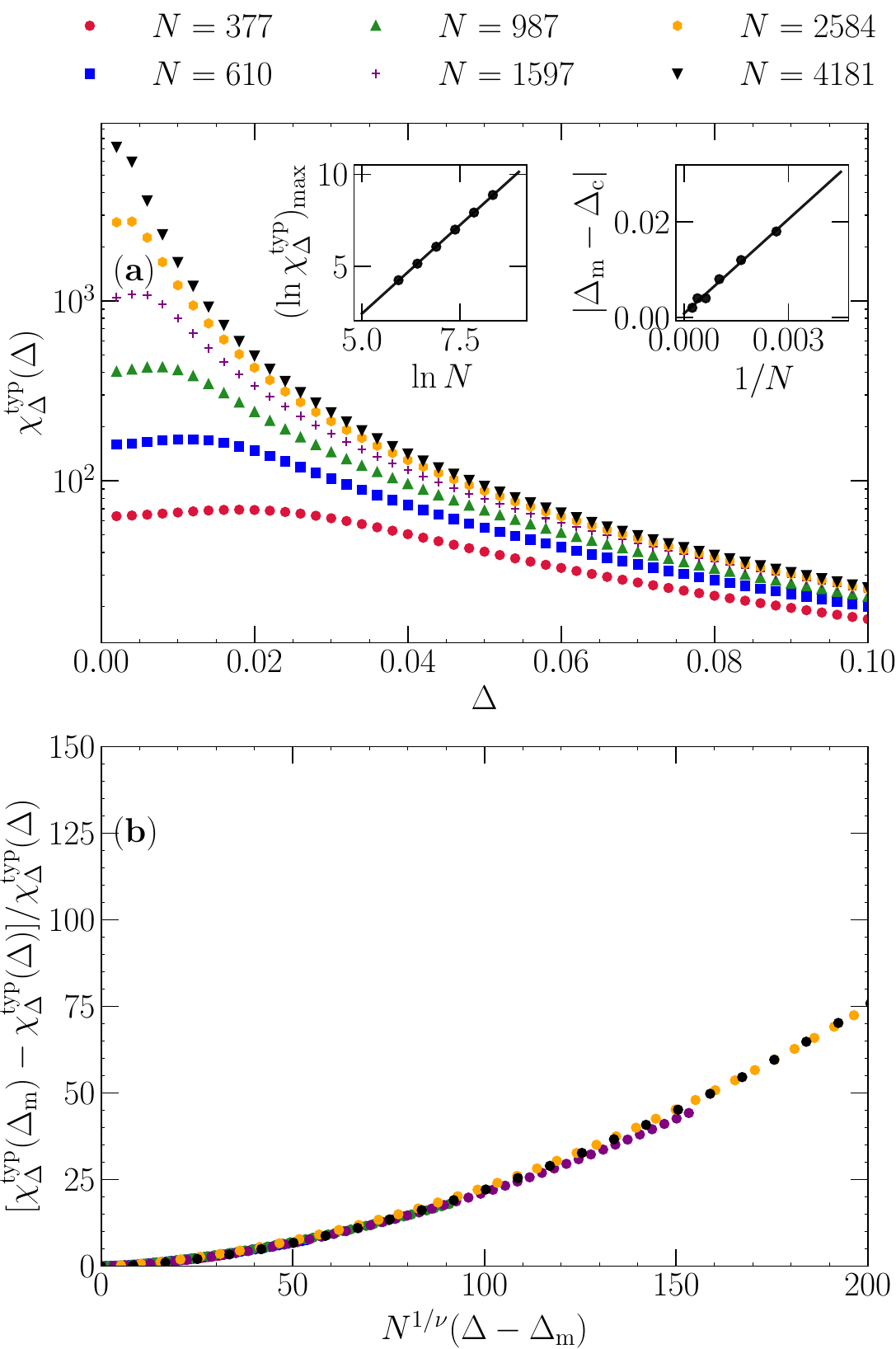} 
\caption{
Scaling properties of the disordered AAH model with Fibonacci potential at $V = 1$: (a) The quantum fidelity susceptibility $\chi_\Delta^{\rm typ}$ as a function of the Fibonacci potential strength $\Delta$ for various system sizes. The inset shows the log-log scaling of $(\ln \chi_\Delta^{\rm typ})_{\rm max}$ versus system size $N$ and the absolute value of the difference between the pseudocritical point and the critical point $|\Delta_{\rm m} - \Delta_{\rm c}|$ versus $1/N$ with $\Delta_{\rm c} = 0$. (b) The rescaled fidelity susceptibility, $[\chi_\Delta^{\rm typ}(\Delta_{\rm m}) - \chi_\Delta^{\rm typ}(\Delta)] / \chi_\Delta^{\rm typ}(\Delta)$, plotted as a function of the scaling variable $N^{1/\nu}(\Delta - \Delta_{\rm m})$ for $\nu = 1$.
}
\label{fsfd}
\end{figure}

\begin{figure}[!htb]
\centering
\includegraphics[width=0.45\textwidth]{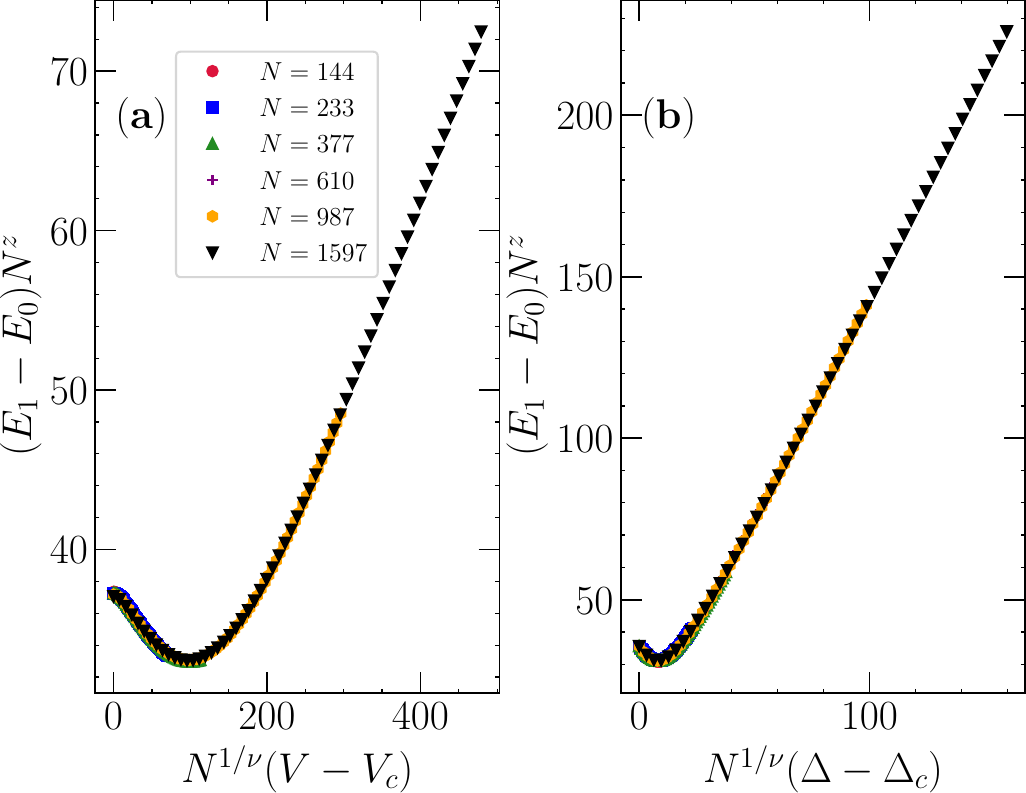} 
\caption{
The Scaling behavior of the gap in the disordered AAH model with Fibonacci potential: (a) The rescaled gap, $(E_1 - E_0) N^z$, as a function of $N^{1/\nu} (V - V_c)$ for $\Delta = 0.1$. (b) The rescaled gap, $(E_1 - E_0) N^z$, as a function of $N^{1/\nu} (\Delta - \Delta_c)$ for $V = 1$.
The results of both data-collapse analyses indicate that the dynamical critical exponent $z$ is equal to 2.
} 
\label{gap}
\end{figure}

\begin{figure}[!htb]
\centering
\includegraphics[width=0.45\textwidth]{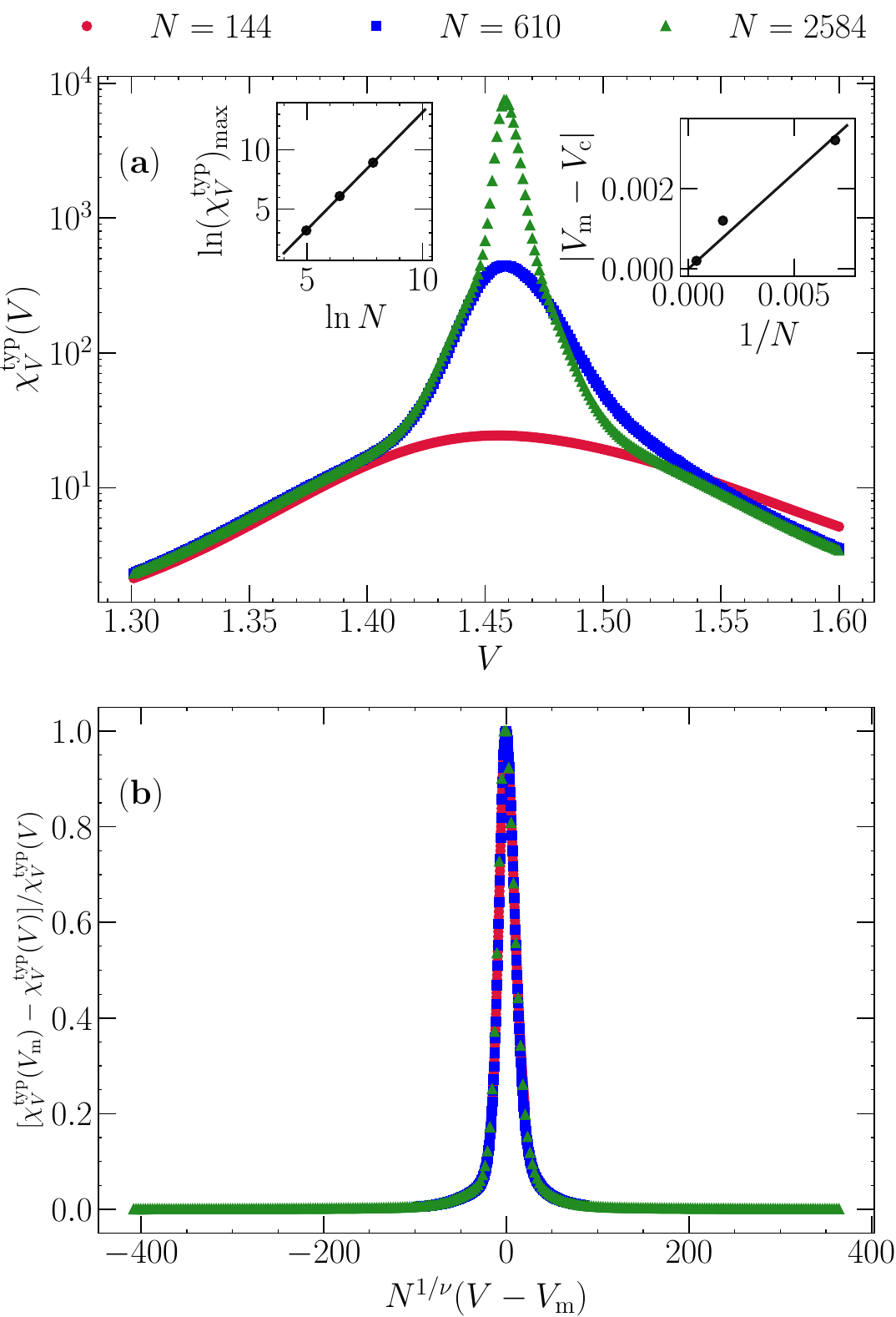} 
\caption{
Scaling properties of the AAH model with the staggered potential for $\Delta = 0.1$: (a) The quantum fidelity susceptibility, $\ln \chi_V^{\rm typ}$, as a function of the AAH potential strength $V$ at $\Delta = 0.1$ for various system sizes. The inset shows the log-log scaling of $(\ln \chi_V^{\rm typ})_{\rm max}$ versus system size $N$ and the absolute value of the difference between the pseudocritical point and the critical point $|V_{\rm m} - V_{\rm c}|$ versus $1/N$ with $V_c=1.4592$. (b) The rescaled fidelity susceptibility, $[\chi_V^{\rm typ}(V_{\rm m}) - \chi_V^{\rm typ}(V)] / \chi_V^{\rm typ}(V)$, is plotted as a function of the scaling variable $N^{1/\nu}(V - V_{\rm m})$ for $\nu = 1$.
}
\label{fsstag}
\end{figure}

\section{Fidelity susceptibility and Scaling behavior in the AAH model with Fibonacci potential}
\label{sec4}

In previous studies, fidelity susceptibility has been demonstrated to be a powerful tool for investigating phase transitions in various disordered systems~\cite{PhysRevB.79.184427,PhysRevA.99.042117,PhysRevA.105.013315,PhysRevB.106.144205,PhysRevE.109.054123}. 
Building on this foundation, we employ fidelity susceptibility to identify critical points, which is defined as~\cite{PhysRevE.76.022101}:
\begin{eqnarray}
\chi_\lambda = \sum_{m>0} \frac{\left|\langle \Psi_m |\partial_{\lambda}\hat{H}| \Psi_0 \rangle\right|^2}{(E_m - E_0)^2}, 
\label{generalizedsusceptibility}
\end{eqnarray}
$|\Psi_m \rangle$  is the $m$-th eigenstate of the system ($|\Psi_0 \rangle$ is the ground-state wavefunction), $E_m(E_0)$ is the corresponding eigenenergy.
To address the large fluctuations induced by disorder, we calculate the typical value of fidelity susceptibility, defined as $\chi_\lambda^{\rm typ} \equiv \exp{\left(\overline{\ln \chi_\lambda}\right)}$, where $\overline{(\cdot)}$ denotes disorder averaging~\cite{PhysRevB.104.195117}.

The maximal value of the typical fidelity susceptibility is expected to obey a power-law relation with the system size, as described in~\cite{you2015generalized}: 
\begin{eqnarray}
(\chi_{\lambda}^{\rm typ})_{\rm max} \sim N^{2/\nu},
\label{equ:scaling1}
\end{eqnarray}
where $\nu$ denotes the correlation-length critical exponent.  
As the system size increases, the pseudocritical points $\lambda_m$ tend to approach the critical points $\lambda_c$, satisfying 
$|\lambda_m - \lambda_c| \propto N^{-1/\nu}$.
The typical fidelity susceptibility of a finite system of size $N$, in the vicinity of a quantum critical point, is expected to obey a universal scaling form~\cite{PhysRevB.81.064418}: 
\begin{eqnarray}
\chi_{\lambda}^{\rm typ}(\lambda) = N^{2/\nu} f_\lambda(N^{1/\nu}|\lambda - \lambda_{\rm m}|), 
\label{equ:scaling3}
\end{eqnarray}
where $f_\lambda$ represents an unknown universal scaling function for the typical fidelity susceptibility with respect to $\lambda$.

First, we investigate the critical behavior of the disordered AAH model with Fibonacci potential using the typical fidelity susceptibility. The logarithmic scale of typical fidelity susceptibility, $\chi_V^{\rm typ}$, for the disordered AAH model with Fibonacci potential as a function of the AAH potential strength $V$ is presented in Figs.~\ref{fsfv}(a-b). The typical fidelity susceptibility reaches its maximum, $(\chi_V^{\rm typ})_{\rm max}$, at the pseudocritical point $V_{\rm m}$ when the system size is finite. As the system size increases, the peaks of $\chi_V^{\rm typ}$ sharpen, and the pseudocritical point $V_{\rm m}$ converges toward the critical point $V_{\rm c} = 0$, allowing for a reliable linear fit. The corresponding log-log scaling relations for $(\chi_V^{\rm typ})_{\rm max}$ versus system size $N$, and $|V_{\rm m} - V_{\rm c}|$ versus $1/N$ are shown in the insets of Fig.~\ref{fsfv}.  For $\nu = 1$, the rescaled typical fidelity susceptibility, $\left[\chi_V^{\mathrm{typ}}\left(V_{\rm m}\right)-\chi_V^{\mathrm{typ}}(V)\right] / \chi_V^{\mathrm{typ}}(V)$, collapses onto a single curve when plotted against the scaling variable $N^{1/\nu}(V - V_{\rm m})$, as demonstrated in Fig.~\ref{fsfv}(b).
This behavior validates the estimated critical parameter and confirms the applicability of the single-parameter scaling hypothesis described in Eq.~(\ref{equ:scaling3}).

Subsequently, the typical fidelity susceptibility for the Fibonacci potential $\Delta$ are presented in Figs.~\ref{fsfd}(a-b) for various system sizes with $V = 1$. Figure~\ref{fsfd}(a) and its insets show that $(\chi_\Delta^{\rm typ})_{\rm max} \propto N^{2/\nu}$ at $\Delta = \Delta_{\rm m}$, and the pseudocritical point $\Delta_{\rm m}$ approaches the critical point  $\Delta_{\rm c}$, satisfying $|\Delta_{\rm m} - \Delta_{\rm c}| \propto 1/N$ with $\Delta_{\rm c} = 0$. The critical exponent $\nu$ remains 1, as indicated by the data collapse shown in Fig.~\ref{fsfd}(b).

Besides, the gap between the ground-state and the first-excited state obey the following scaling behavior~\cite{bu2022quantum} 
\begin{eqnarray}
E_1-E_0=N^{-z}g_\lambda(N^{1/\nu}|\lambda-\lambda_c|),
\end{eqnarray}
where $E_0$ ($E_1$) is the groundstate (first-excited state) energy, $z$ is the dynamical critical exponent, $g_\lambda$ represents a unknown regular universal scaling function for the gap with respect to $\lambda$, and $\lambda=V,\Delta$.
The rescaled gap $(E_1-E_0)N^z$ is presented as a function of the proper scaling variable  $N^{1/\nu}(V-V_{c})$ in Fig.~\ref{gap}(a). The corresponding result for the rescaled gap $(E_1-E_0)N^z$ with respect to $N^{1/\nu}(\Delta-\Delta_{\rm c})$ is shown in Fig.~\ref{gap}(b). In both cases shown in Figs.~\ref{gap}(a-b), the rescaled curves exhibit excellent collapse for $\nu = 1$ and $z = 2$. This behavior contrasts with that of the AAH model without disorder ($\nu = 1$, $z = 2.375$) and the Anderson model ($\nu = 2/3$, $z = 2$)~\cite{PhysRevA.99.042117}.

We also investigate the scaling behavior of the disordered AAH model with the staggered potential. The typical fidelity susceptibility, $\chi_V^{\rm typ}$, as a function of the AAH potential strength $V$ with $\Delta = 0.1$, is shown in Figs.~\ref{fsstag}(a-b). The maximum value of $\chi_V^{\rm typ}$, denoted as $(\chi_V^{\rm typ})_{\rm max}$, occurs at $V = V_{\rm m}$ and follows the scaling relation $(\chi_V^{\rm typ})_{\rm max} \propto N^{2/\nu}$. As the system size $N$ increases, the pseudocritical point $V_{\rm m}$ approaches the critical point $V_{\rm c}=1.4592$, satisfying $|V_{\rm m} - V_{\rm c}| \propto 1/N$, as shown in the insets of Fig.~\ref{fsstag}(a). In Fig.~\ref{fsstag}(b), the rescaled typical fidelity susceptibility, $[\chi_V^{\rm typ}(V_{\rm m}) - \chi_V^{\rm typ}(V)] / \chi_V^{\rm typ}(V)$, is plotted as a function of the scaling variable $N^{1/\nu}(V - V_{\rm m})$. Near the critical point $V_{\rm m}$, the curves for different system sizes collapse onto a single universal curve for $\nu = 1$. 
We find that the disordered AAH models with staggered and Fibonacci potentials exhibit the same correlation-length critical exponent despite the differences in localization behavior.

\section{Summary}
\label{sec5}
In this paper, we investigate the quantum criticality and localization transition in the ground state of the disordered  Aubry-Andr\'{e}-Harper (AAH) model with different discrete potentials, such as staggered, Fibonacci, and Thue-Morse potentials. The phase diagram for each model consists of three distinct phases: extended, critical, and localized. When the staggered potential is introduced, a localization transition is observed as the AAH potential is tuned. However, when the Fibonacci or Thue-Morse potentials are applied, the system immediately becomes localized. Additionally, the inverse participation ratio (IPR) for different models follows the complexity of the corresponding discrete potential when the amplitude of the AAH potential is weak. As the AAH potential becomes dominant, this trend is disrupted. Furthermore, we focus on the critical behaviors of the disordered AAH model with the Fibonacci potential. 
Through the analysis of the typical fidelity susceptibility and the energy gap, we determine that \(\nu = 1\) and \(z = 2\). The former is consistent with the results of the AAH model without disorder (\(\nu = 1, z = 2.375\)), while the latter agrees with those of the Anderson model (\(\nu = 2/3, z = 2\))~\cite{PhysRevA.99.042117}.
We observe that, despite the differences in localization behavior, the disordered AAH models with staggered and Fibonacci potentials display the same correlation-length critical exponent.

The study of hybrid models is essential for advancing our understanding of nontrivial quantum systems. The individual components of hybrid models have already been independently realized in experiments; for instance, the AAH model has been successfully implemented in cold-atom optical lattice experiments~\cite{billy2008direct}, while discrete potentials have been realized using quantum walk techniques~\cite{PhysRevLett.93.180601, PhysRevA.93.032329, nguyen2020quantum}. Previous studies have demonstrated the feasibility of implementing the AAH model in circuit systems~\cite{ganguly2023electrical}, where localization transitions and mobility edges have been observed. Since introducing additional disorder is relatively straightforward in circuit setups, these systems present a promising platform for experimentally validating the findings of this work.  Furthermore, the experimental determination of critical exponents, such as the correlation length exponent \(\nu\) and the dynamical critical exponent \(z\), has become increasingly feasible. For example, \(z\) dictates the low-temperature specific heat scaling \(C_v \sim T^{-z}\), which can be inferred from the density of states \(\rho \sim \omega^{1/z-1}\) or through the Kibble-Zurek mechanism~\cite{PhysRevB.99.094203}. 
The advent of hybrid potential models presents novel opportunities to explore universality in quasiperiodic systems. Consequently, investigating the critical exponents of models with hybrid potentials is of significant scientific interest.

\begin{acknowledgments}
The authors appreciate very insightful discussions with
Xingbo Wei. 
This work is supported by the National Natural Science Foundation of China (NSFC) under Grant Nos.~12404285, 12174194, 12474492, 12461160324, the Zhejiang Provincial Natural Science Foundation of China under Grant No.~LQN25A040003, and the Science Foundation of Zhejiang Sci-Tech University under Grant No.~23062182-Y. 
\end{acknowledgments}

\section*{DATA AVAILABILITY}
The data that support the 
findings of this article are openly available \cite{Yi2025Github}.
\bibliography{refs}

\clearpage
\widetext
\setcounter{equation}{0}
\setcounter{figure}{0}
\setcounter{table}{0}
\setcounter{section}{0}
\setcounter{tocdepth}{0}
 \renewcommand{\thefigure}{S\arabic{figure}}
\numberwithin{equation}{section} 
\begin{center}
{\bf \large Supplemental Material for “Unveiling quantum criticality of disordered Aubry-Andr\'{e}-Harper model via typical fidelity susceptibility” }
\end{center}

\section{Parity Structure of Fibonacci Numbers and Choice of System Size}
\label{app_fibonachi}

In numerical calculations, the golden ratio $\alpha = \frac{\sqrt{5} - 1}{2}$ is approximated by the rational form $\alpha = \frac{F_k}{F_{k+1}}$, where $F_k$ is the $k$-th Fibonacci number. Correspondingly, the total number of lattice sites is chosen as $N = F_{k+1}$. Here, $F_k$ denotes the $k$-th Fibonacci number, defined recursively by $F_{k+1} = F_k + F_{k-1}$ with initial conditions $F_0 = F_1 = 1$. Notably, the Fibonacci sequence $F_k$ exhibits a three-periodic parity pattern:
$\{ F_{3l+1} \mid \ldots, 55, 233, 987, \ldots \}$,
$\{ F_{3l+2} \mid \ldots, 89, 377, 1597, \ldots \}$,
$\{ F_{3l+3} \mid \ldots, 144, 610, 2584, \ldots \}$,
where $l$ is an integer. This pattern indicates that every third Fibonacci number is even, while the others are odd. To minimize boundary effects and maintain compatibility with the staggered potential in the disordered AAH model, we choose $N = 144, 610, 2584$, which are both Fibonacci numbers and even numbers.

\section{Finite-Size Effects on the IPR in Disordered AAH Models with Different Potentials}

\begin{figure}[!h]
\centering
\includegraphics[width=0.95\textwidth]{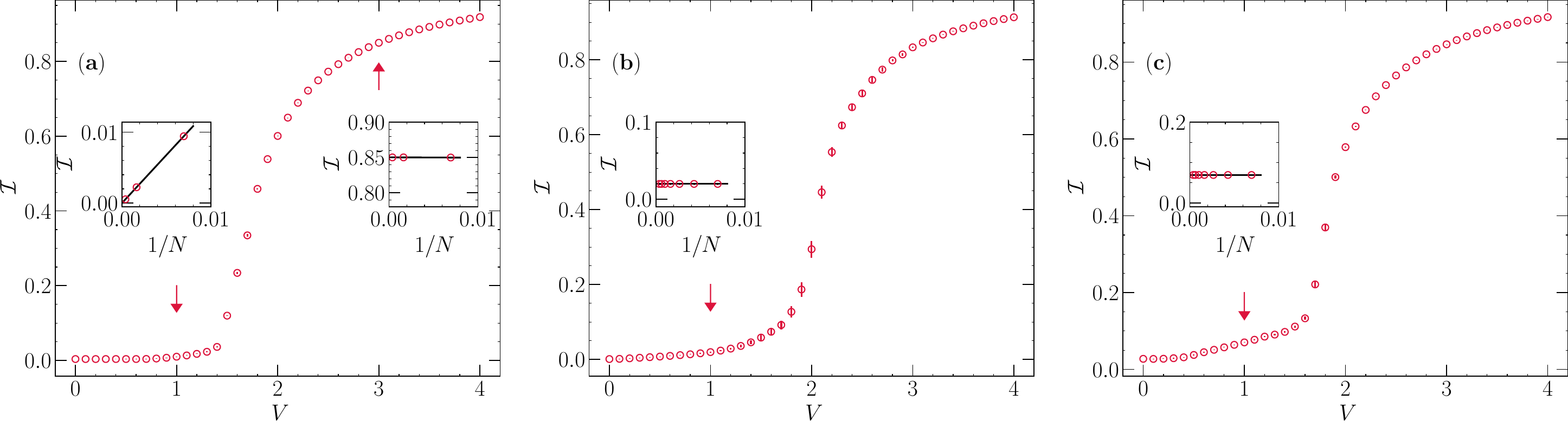}
\caption{
The IPR as a function of $V$ for the disordered AAH models with (a) staggered potential, (b) Fibonacci potential, and (c) Thue-Morse potential, when $N = 987$ and $\Delta = 0.1$, is shown. The corresponding insets display the IPR as a function of $N$ for the disordered AAH models with various potentials.
}   
\label{ipr_sft}
\end{figure}

We present the variation of the inverse participation ratio (IPR) with respect to the AAH potential strength $V$ in the disordered AAH model in Figs.~\ref{ipr_sft}(a-c). The overall trend across the three figures indicates that as $V$ increases, the IPR also increases. However, only the AAH model with the staggered potential exhibits a phase transition from the extended phase to the localized phase when the AAH potential is finite (see Fig.~\ref{ipr_sft}(a)). Insets in Fig.~\ref{ipr_sft}(a) illustrate the variation of the IPR with respect to $N$ when $V = 1.0$ and $3.0$. For $V = 1.0$, the IPR is proportional to $1/N$, signaling that the system is in the extended phase. For $V = 3.0$, the IPR is proportional to $N^{0}$, indicating that the system has entered the localized phase. The IPR as a function of $V$ for the disordered AAH models with Fibonacci and Thue-Morse potentials is shown in Figs.~\ref{ipr_sft}(b) and \ref{ipr_sft}(c). For $\Delta > 0$, the system becomes immediately localized. The insets in Figs.~\ref{ipr_sft}(b) and \ref{ipr_sft}(c) display IPR $\propto N^{0}$ when $V = 1$, which is a characteristic signature of the localized state.

\section{IPR $\mathcal{I}$ versus $V$ for Disordered AAH Models with Various Potentials at $\Delta = 0.1,\ 0.3,\ 0.5,\ 0.7,\ 0.9$}

\begin{figure}[!t]
\centering
\includegraphics[width=0.45\textwidth]{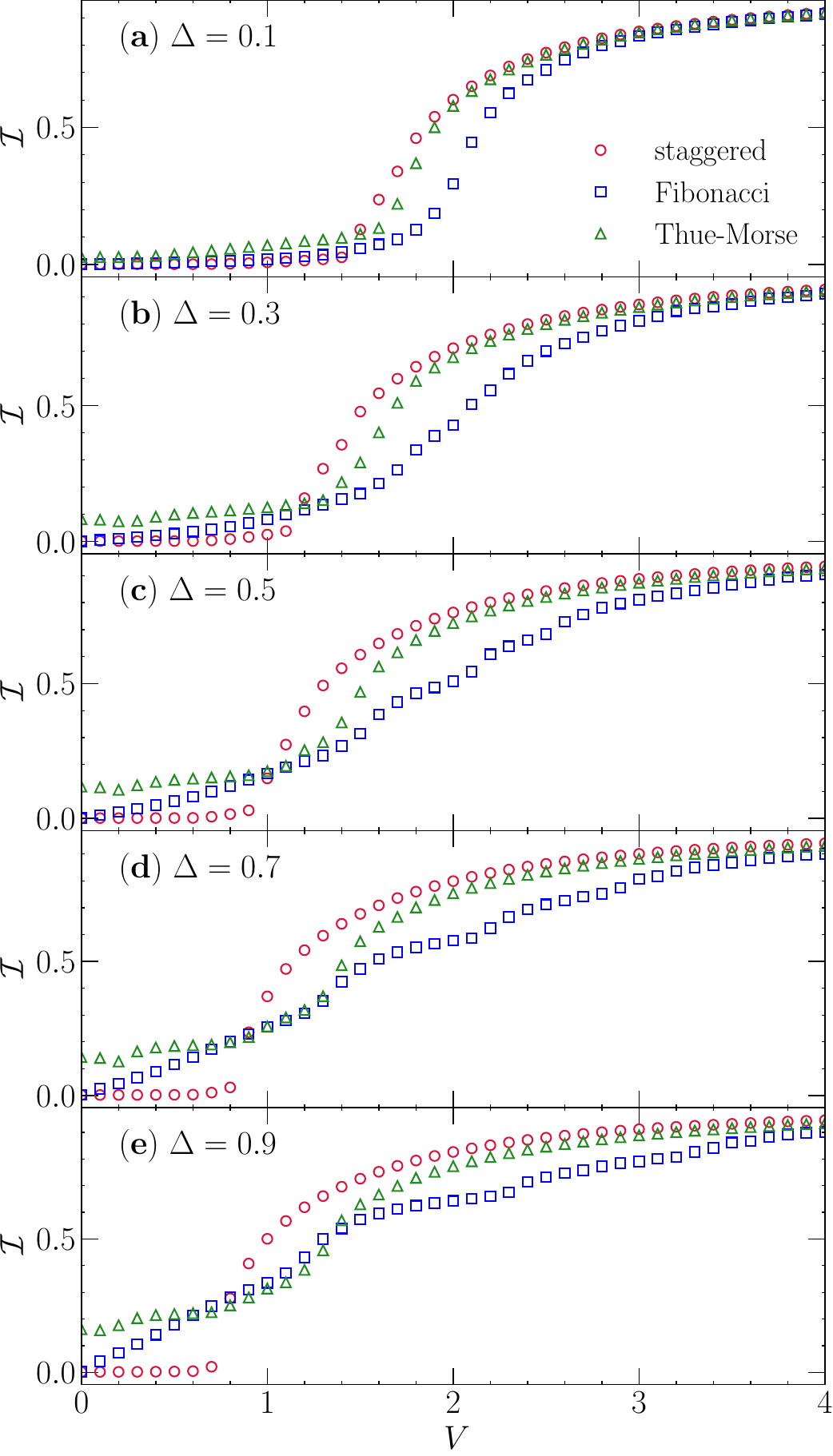}
\caption{
The IPR ${\cal I}$ with respect to $V$ for the disordered AAH models with various potentials for $N=987$,  $\Delta = 0.1, 0.3, 0.5, 0.7, 0.9$.
}   
\label{ipr_Delta13579}
\end{figure}

Figures~\ref{ipr_Delta13579}(a-e) display the inverse participation ratio (IPR) ${\cal I}$ as a function of $V$ for the disordered AAH models with different potentials, for values of $\Delta = 0.1, 0.3, 0.5, 0.7, 0.9$. The amplitude of the IPR ${\cal I}$ as a function of $V$ for the disordered AAH models with various potentials follows the order of complexity of the corresponding sequences, regardless of the value of $\Delta$.

\end{document}